\documentstyle[12pt,lnfprep,epsfig]{article}

\newcommand{\beq}{\begin{equation}}
\newcommand{\eeq}{\end{equation}}
\newcommand{\ba}{\begin{array}}
\newcommand{\ea}{\end{array}}
\newcommand{\beqa}{\begin{eqnarray}}
\newcommand{\eeqa}{\end{eqnarray}}

\newcommand{\no}{\nonumber}
\newcommand{\lsim}{\stackrel{<}{_\sim}}

\newcommand{\lt}{<}
\newcommand{\Imm}{\mbox{Im}}
\newcommand{\Real}{\mbox{Re}}
\newcommand{\Ko}{{K}^0}
\newcommand{\Kob}{\bar{K}^0}

\newcommand{\eps}{\epsilon}
\newcommand{\epsp}{\epsilon'}

\newcommand{\Kol}{$K_{L}$}
\newcommand{\Kos}{$K_{S}$}
\newcommand{\lum}{cm$^{-2}$s$^{-1}$}
\newcommand{\ilum}{cm$^{-2}$}

\newcommand{\PL}[3]{{Phys. Lett.}       {\bf #1} {(19#2)} {#3}}

\newcommand{\PRL}[3]{{Phys. Rev. Lett.} {\bf #1} {(19#2)} {#3}}
\newcommand{\PR}[3]{{Phys. Rev.}        {\bf #1} {(19#2)} {#3}}
\newcommand{\NP}[3]{{Nucl. Phys.}       {\bf #1} {(19#2)} {#3}}
\newcommand{\ZP}[3]{{Z.  Phys.}         {\bf #1} {(19#2)} {#3}}

\newcommand{\RMP}[3]{{Rev. Mod. Phys.}  {\bf #1} {(19#2)} {#3}}

\newcommand{\Header}{
  \begin{tabular}{rl}
  \hspace{-.4cm}\includegraphics{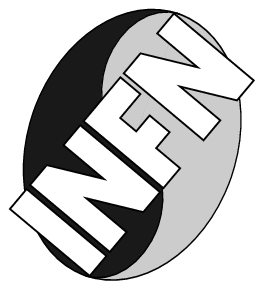} &
    \renewcommand{\arraystretch}{0.5}
    \begin{tabular}{r}
      {\hspace{1cm}~\LARGE\sffamily LABORATORI~ NAZIONALI~ DI~ FRASCATI}\\
      \\
      {\Large\sffamily SIS-Pubblicazioni}\\
    \end{tabular}
    \renewcommand{\arraystretch}{1}
  \end{tabular}
  \vskip 1cm
  \begin{flushright}
  \renewcommand{\arraystretch}{0.5}
    \begin{tabular}{r}
      {\underline{LNF--98/004(P)}}\\ 
      {\small February 1998} \\    
      \\       
      {\small\tt  hep-ph/9802345}\\        
    \end{tabular}
  \end{flushright}
  \renewcommand{\arraystretch}{1}
  \vspace{2cm} 
  }
\begin{document}
\begin{titlepage}
\title{ 
  \Header
  {\large \bf Searching for $K_L\to\pi^0\nu\bar{\nu}$ at a $\Phi$--factory } \\
\author{F. Bossi, G. Colangelo and G. Isidori \\
\\
  {\it INFN, Laboratori Nazionali di Frascati,} \\ {\it I--00044 Frascati, Italy}
} }
\maketitle
\baselineskip=14pt
\begin{abstract}
The perspectives of a search for the rare decay
$K_L\to\pi^0\nu\bar{\nu}$ at a $\Phi$--factory are discussed.  
After a general analysis, we focus on the realistic
case of KLOE and DA$\Phi$NE, showing that limits of the order
of $10^{-9}$ on BR($K_L\to\pi^0\nu\bar{\nu}$)
are achievable in the next few years.
We also discuss the theoretical implications of this kind of measurements.
\end{abstract}

\vspace*{\stretch{2}}
\begin{flushleft}
\vskip 2cm
{PACS: 12.15.-y,~13.20.Eb} 
\end{flushleft}

\end{titlepage}
\pagestyle{plain}
\baselineskip=17pt
\pagenumbering{arabic}

\section{Introduction}   

Flavor--changing neutral--current
kaon decays provide a fundamental pro\-be to investigate the flavor 
structure of electroweak interactions \cite{GIM,GL}. 
Among them, $K\to\pi\nu\bar{\nu}$ transitions 
can be considered the ``gold--plated'' channels 
because of their freedom from long--distance uncertainties
\cite{Litt,BB,BB2mt,LW,MP}. A measurement of the $K\to\pi\nu\bar{\nu}$
decay widths would provide unique informations on fundamental parameters of 
the Standard Model, and possibly also on the physics beyond it, as  has
been recently emphasized in \cite{GN,Burd,NW,BRS}.
In recent years an important experimental 
challenge has been undertaken to observe such transitions,
and very recently preliminary evidence for $K^+\to\pi^+\nu\bar{\nu}$ 
was found \cite{BNL}. Despite this success, the 
experimental difficulties in the neutral channels 
($K_{L,S}\to\pi^0\nu\bar{\nu}$) are still far form being solved.

Although the signature of the $K_L\to\pi^0\nu\bar{\nu}$ decay looks at
first straightforward (two photons whose invariant mass equals that of 
a $\pi^{0}$ and nothing else), the problem of backgrounds rejection has
so far proven to be very difficult to handle, resulting in 
rather poor limits on the corresponding branching ratio. 
In fact, the decay channels of the $K^{0}_{L}$ into 2 or 3 $\pi^{0}$'s
have branching ratios several orders of magnitude larger than the one
expected for the signal, requiring, therefore, a very high photon detection 
capability. This is particularly important 
in view of the practical impossibility to completely reconstruct the decay 
kinematics at hadron machines, where all the searches for
$K_L\to\pi^0\nu\bar{\nu}$ have been performed thus far. 
Moreover, at these machines, kaon beams are accompanied by unwanted
neutral--hadron halos, which can fake the signal either by interaction
with the residual gas in the decay volume or via decays such as
$\Lambda\to\pi^0 n$.  

With the present paper we want to draw  attention to the fact that
many of the problems listed above have a natural solution if the 
search is performed at a $\phi$--factory. 
The \Kol\ beam available at a $\phi$--factory is 
{\it monochromatic}, which allows the complete reconstruction of the decay
kinematics, greatly helping in the rejection of the 
most dangerous physics background i.e. $K_L\to\pi^0\pi^0$. 
Moreover,  since it is  a {\it tagged} beam, it is also free from the
background due to accidentals which can mimic the signal.  

The paper is organized as follows. In the next section we briefly introduce 
the theoretical framework needed to describe this decay, and discuss the
implications of a measurement of the  $K\to\pi\nu\bar{\nu}$
decay widths. In section 3 we describe the present experimental
status on $K_L\to\pi^0 \nu\bar{\nu}$ searches and the prospects for future 
measurements. Section 4 is devoted to the study of the feasibility of 
this measurement at a $\phi$--factory, with special attention to 
what can be obtained, in a short time frame, at facilities 
which are at present in the commissioning phase. Our conclusion are 
then summarized in the final section. 

\section{Theoretical overview}
Within the Standard Model, $K\to\pi\nu\bar{\nu}$
transitions can be described by means of the 
following effective four--fermion Hamiltonian
\beq
{\cal H}_{\rm eff} = \frac{\alpha G_F }{2\sqrt{2}\pi\sin^2\Theta_W}
\sum_{l=e,\mu,\tau} C^l~ 
{\bar s}\gamma^\mu (1-\gamma_5)
d~\bar{\nu}_l\gamma_\mu(1-\gamma_5)\nu_l ~+~ \mbox{h.c.}
\label{Heff}
\eeq 
The Wilson Coefficients $C^l$ have been calculated by Buchalla and
Buras including  next--to--leading order QCD corrections \cite{BB}
and, recently, also $O(G_F^2m_t^4)$ effects \cite{BB2mt}. 
Neglecting the latter,  which represent at most a few percent 
correction, we can write  \cite{BB}
\beq
C^l = \lambda_c X^l_{NL} + \lambda_t  X(m^2_t/M_W^2) \; , 
\label{Cl}
\eeq
where $\lambda_q=V_{qs}^* V_{qd}~$,~$V_{ij}$ denotes the 
Cabibbo--Kobayashi--Maskawa (CKM) \cite{CKM}
matrix elements and the $X$ functions can be found in \cite{BB}
(see also \cite{BBL}). Numerically, $X^l_{NL} \sim 10^{-3}$ and
$X(m^2_t/M_W^2) \simeq 1.5$;
thus  charm and top contributions to the real
part of $C^l$ are comparable since
$\mbox{Re}(\lambda_c)/\mbox{Re}(\lambda_t) \sim {\cal O}(10^{-3})$,
while the top contribution dominates the imaginary part because
$\mbox{Im}(\lambda_c)/\mbox{Im}(\lambda_t) \sim {\cal O}(1)$.
 
The matrix elements of the Hamiltonian (\ref{Heff}) between kaon and pion
states are well known since they are related by isospin symmetry to those
relevant for the corresponding (charged current) semileptonic
decays. Neglecting isospin breaking effects we can write for 
the case of our interest 
\beqa
&& \left|\sqrt{2}\langle\pi^0|{\bar s}\gamma^\mu d|\Ko  \rangle\right| =
   \left|\sqrt{2}\langle\pi^0|{\bar d}\gamma^\mu s|\Kob \rangle\right| =\no\\
&& \langle \pi^+(p_\pi) | {\bar s}\gamma^\mu d | K^+(p_K) \rangle = 
f_+(q^2) (p_\pi^\mu+p_K^\mu) + {\cal O}(q^\mu=p_K^\mu-p_\pi^\mu)~,
\label{melem}
\eeqa
where 
\beq
f_+(q^2) = 1+ \lambda { q^2 \over M^2_{\pi^+} }\qquad  \mbox{and} \qquad
 \lambda = (0.030 \pm 0.002)~.
\eeq
Thus the three decay modes have the same spectrum and only differ 
by a normalization factor. In the charged case we find 
\beq
\frac{\mbox{d}\Gamma(K^+\to\pi^+\nu_l\bar{\nu}_l)}{\mbox{d}E_\pi}
= \frac{ \alpha^2 G_F^2 | C^l |^2 M_K }{ 48 \pi^5 \sin^4
  \Theta_W}\left|f_+(q^2)\right|^2 
(E^2_\pi - M^2_\pi)^{3/2}~,
\eeq
where 
\beq\label{energyrange}
M_\pi \leq E_\pi \leq E_\pi^{\rm max} = 
 \frac{M_K}{2}\left(1+\frac{M_\pi^2}{M_K^2}\right)\qquad
\mbox{and}\qquad q^2=M_K^2+M_\pi^2-2M_KE_\pi~.
\eeq

The relative phase between the neutral matrix elements in 
(\ref{melem}) depends on the phase convention for 
$|\Ko \rangle$ and $|\Kob\rangle$ states. 
Assuming the matrix elements to be real 
and imposing $CP|\Ko \rangle=|\Kob\rangle$ 
leads to $\langle\pi^0|{\bar s}\gamma^\mu d|\Ko  \rangle$ =
$\langle\pi^0|{\bar d}\gamma^\mu s|\Kob \rangle$.
Then defining as usual 
\beq
|K_{L,S}\rangle = {1 \over \sqrt{2(1+|\eps|^2)}}\left(
(1+\eps) |\Ko  \rangle \mp (1-\eps)  |\Kob\rangle \right)
\eeq
and neglecting the suppressed ${\cal O}(\epsilon)$ 
terms\footnote{~In our phase convention 
$\Imm \; \eps \sim \Real \; \eps \sim {\cal O}(10^{-3})$.} leads to 
\beqa
\frac{ A(K_S\to \pi^0\nu_l\bar{\nu}_l) }{ |A(K^+\to\pi^+\nu_l\bar{\nu}_l)| } 
&=& \frac{\Real C^l}{|C^l|} ~\simeq~ \frac{\rho^l_0 -\bar\rho}{
\sqrt{(\sigma\bar\eta)^2 +(\bar\rho- \rho_0^l)^2} }~, \label{rap1}\\ 
\frac{ A(K_L\to \pi^0\nu_l\bar{\nu}_l) }{ |A(K^+\to\pi^+\nu_l\bar{\nu}_l)| }
&=& \frac{i\Imm C^l}{|C^l|} ~\simeq~ -\frac{ i\sigma\bar\eta }{
\sqrt{(\sigma\bar\eta)^2 +(\bar\rho- \rho_0^l)^2} }~. \label{rap2} 
\eeqa
Here we have used the Wolfenstein parametrization of
the CKM matrix \cite{Wolf} in its modified version (first
introduced by \cite{SS} and then redefined in \cite{BLO}):
\beqa
\lambda_t&=& -\sigma^{-1/2} A^2 \lambda^5 (1-\bar{\rho}-i \sigma
\bar{\eta}) \doteq |\lambda_t| e^{i \beta}
\; \; , \nonumber \\
\mbox{Re}(\lambda_c)&=&-\lambda \sigma^{-1/2}\; , \nonumber \\
\mbox{Im}(\lambda_c)&=& -\mbox{Im} (\lambda_t) \; ,
\eeqa
where\footnote{~For simplicity we use a definition of $\lambda,~ \rho$ and
  $\eta$ which is not exactly that given in \cite{BLO}. On the other hand,
  the relative difference is of ${\cal O}(\lambda^6)$ 
  which is far beyond the accuracy we need.} 
$\lambda = |V_{us}|$, $(\rho+i\eta)=V_{ub}^*/( V_{us} V_{cb})$,
$\sigma^{-1/2} = (1-\lambda^2/2)$, $\bar\rho=\sigma^{-1/2}\rho$ and
$\bar\eta=\sigma^{-1/2}\eta$. 
The dominant contribution to all the amplitudes 
is independent of the lepton flavour and is
proportional to the $\lambda_t$ term in 
(\ref{Cl}).
The charm contamination is totally negligible in the $K_L$ decay, but
induces the largest theoretical uncertainty in the evaluation of the real
part of $C^l$. This contribution is parametrized by 
\beq
\rho^l_0 -1 = \frac{X^l_{NL}}{A^2\lambda^4 X(m^2_t/M_W^2)} \lsim 0.3  
\quad\protect\cite{BBL}~.
\eeq
For later convenience we also recall that the latest numerical analysis of
the CKM matrix yields \cite{BBL}:
\beq
\lambda= 0.2205 \pm 0.0018 \; , ~~~|\lambda_t| = (3.5 \pm 0.5 ) \times 10^{-4}
\; ,~~~\beta \sim (10 \div 30)^\circ \; \;. 
\eeq

We remark that since the $\pi^0\nu_l\bar{\nu}_l$ state produced by 
${\cal H}_{\rm eff}$ is $CP$--even, the
$K_L\to \pi^0\nu_l\bar{\nu}_l$ amplitude has to 
vanish in the limit of exact $CP$ symmetry, as it is apparent from
Eq. (\ref{rap2}).

In principle there is also a long--distance contribution, generated 
by light quark rescattering, that can be calculated 
in the framework of Chiral Perturbation Theory. This amounts to a 
few percent correction to $\rho_0^l$ \cite{LW} and,
being much smaller than the scale uncertainty of the charm
contribution, can be safely neglected. In passing, we note that 
long--distance effects vanish at $O(p^2)$ only if exact nonet symmetry is 
assumed, as correctly stated in \cite{LW} (and in contrast to what has been 
claimed in \cite{Geng}). 

Eqs.~(\ref{rap1}-\ref{rap2}) imply an interesting 
relation among the three the decay 
widths\footnote{~When the lepton flavor is not explicitly 
indicated, the sum over neutrino's families is understood.}
\beq
\Gamma(K_L\to\pi^0 \bar{\nu}\nu) + 
\Gamma(K_S\to\pi^0 \bar{\nu}\nu)  = \Gamma(K^+ \to\pi^+
\bar{\nu}\nu)~.
\label{tri}
\eeq
This is a direct consequence of (\ref{melem}) and indeed receives
small corrections due to isospin--breaking terms, which have been evaluated 
in \cite{MP}. These are generated by the mass difference $m_d-m_u$ 
and by electromagnetic effects.

The expressions for the branching ratios of the three decay transitions $K
\to \pi \nu \bar{\nu}$ are as follows:
\beqa
BR(K^+ \to \pi^+ \nu \bar{\nu}) &=& \kappa_+ {1 \over 3} \sum_l
\left|C^l/\lambda^5 \right|^2 \;\; ,  
\nonumber \\
 BR(K_L \to \pi^0 \nu \bar{\nu}) &=& \kappa_L {1 \over 3} \sum_l \left[
   \mbox{Im}(C^l/\lambda^5)\right]^2 \; \; ,
\\
 BR(K_S \to \pi^0 \nu \bar{\nu}) &=& \kappa_S {1 \over 3} \sum_l \left[
   \mbox{Re}(C^l/\lambda^5) \right]^2 \; \; ,
\nonumber \\
\label{BR}
\eeqa
where
\beq
\kappa_+ = r_{K^+} \frac{ 3 \alpha^2 BR(K^+ \to \pi^0 e^+ \nu) }{2 \pi^2
  \sin^4 \Theta_W } \lambda^8 = 4.11 \times 10^{-11} \; \; .
\eeq
This number has been obtained using $\alpha=1/129$~, $\sin^2 \Theta_W =
0.23$, $BR(K^+ \to \pi^0 e^+ \nu) = 4.82 \times 10^{-2}$, as in
\cite{BBL}, and $r_{K^+}=0.9$ which summarizes isospin breaking corrections
\cite{MP}. The factor $\kappa_+$ gives the order of magnitude one
should expect for $BR(K^+ \to \pi^+ \nu \bar{\nu})$, 
 since $C^l/\lambda^5$ is roughly a
number of order one within the 
Standard Model. A detailed numerical analysis for this last term 
using present constraints on the CKM matrix leads to \cite{BBL}:
\beq
BR(K^+ \to \pi^+ \nu \bar{\nu})_{\rm{SM}} = (8.0 \pm 1.5) \times 10^{-11}
\; \; . 
\eeq
The corresponding $\kappa$ factors for the neutral kaons are defined as
\beq
\kappa_{L,S} = \kappa_+ { \tau_{_{K_{L,S}}} \over \tau_{_{K^+}}} { r_{K^0}
  \over r_{K^+}} \; \; ,
\eeq
where $r_{K^0}=0.94$ has been calculated in \cite{MP}, 
and yields \cite{BBL}: 
\beq
BR(K_L \to \pi^0 \nu \bar{\nu})_{\rm{SM}} = (2.6 \pm 0.9) \times 10^{-11}
\; \; ,
\eeq
while for the $K_S$ the suppression due to the very short lifetime
leads to a branching ratio of order $10^{-13}$.

\subsection{$K\to\pi\nu\bar{\nu}$ beyond the Standard Model}

In most New Physics models $K\to\pi\nu\bar{\nu}$
transitions are still described by the effective Hamiltonian 
(\ref{Heff}), with appropriate Wilson coefficients 
$C^l_{NP}\not=C^l_{SM}$. This is the case for ``typical'' 
supersymmetric models, see e.g. \cite{NW,BRS}, 
but also for SM extensions with strong dynamics
at the electroweak scale \cite{Burd}.  Within this framework,
a convenient parameterization of the Wilson coefficient
is given by \cite{BRS} 
\beqa
C^l_{NP} &=& \lambda_c X^l_{NL} + e^{i\theta_K} r_K
\lambda_t  X(m^2_t/M_W^2) \nonumber \\
&=& \lambda_c X^l_{NL} + e^{i(\theta_K+\beta)}
  r_K |\lambda_t|  X(m^2_t/M_W^2)~,
\label{Clbis}
\eeqa
with $r_K$ real and positive and $-\pi <\theta_K< \pi$
(the SM case is recovered for $r_K=1$ and $\theta_K=0$).
In both supersymmetric and strong--dynamics 
scenarios, the natural size of the parameter $r_K$ is 
$0.5\lsim r_K \lsim 2$, implying small deviations 
of $BR(K^+\to\pi^+\nu\bar{\nu})$ from its SM value.
However, even for $r_K\sim 2$ a large enhancement of 
$BR(K_L\to\pi^0\nu\bar{\nu})$ is possible provided the {\it new--physics phase}
$\theta_K$ is such that  $|\theta_K+\beta| \sim \pi/2$. This possibility is 
not particularly likely but, at least in some
supersymmetric scenarios, still not excluded by data in other channels
\cite{NW,BRS}.

Taking a more general point of view, Grossman and Nir \cite{GN}
have shown that the situation is different if one considers models
with non--vanishing neutrino masses and/or lepton--flavor violations.
In this case one can write different kinds of dimension--six
operators, like ${\bar s}d\bar{\nu}_l\nu_l$ or even
${\bar s}d\bar{\nu}_l\nu_m$. Furthermore, 
if lepton flavor is violated $K_L\to\pi^0\nu\bar{\nu}$ can
receive also $CP$--conserving contributions \cite{GN}.

Interestingly enough, in all these cases the
relation (\ref{tri}) is still valid (up to small 
isospin breaking corrections). This is because
any  $s\to d$ two--quark operator 
carries isospin $\Delta I=1/2$ and thus obeys the isospin relation 
  $|\sqrt{2}\langle\pi^0|O_{{\bar s}d} |\Ko \rangle| $ 
= $|\langle \pi^+ | O_{{\bar s}d} | K^+ \rangle| $. 
The only way to avoid this constraint
is to consider a $\Delta I=3/2$ operator, that is at least
dimension nine for the $s\to d\nu\bar{\nu}$ transition.
Neglecting the effect of this presumably much suppressed operator, 
from Eq.~(\ref{tri}) one can derive a model--independent bound \cite{GN} 
on $BR(K_{L,S}\to\pi^0\nu\bar{\nu})$ in terms of the measured
$BR(K^+\to\pi^+\nu\bar{\nu})$ \cite{BNL}
\beqa
BR(K_L\to\pi^0\nu\bar{\nu}) &<& \frac{\tau_{_{K_L}}}{\tau_{_{K^+}}}
 BR(K^+\to\pi^+\nu\bar{\nu})\left[1+{\cal O}\left(\frac{m_u-m_d}{m_s}\right)\right] 
 \lsim 5\times 10^{-9}~,\qquad \\
BR(K_S\to\pi^0\nu\bar{\nu}) &<& \frac{\tau_{_{K_S}}}{\tau_{_{K^+}}}
 BR(K^+\to\pi^+\nu\bar{\nu})\left[1+{\cal O}\left(\frac{m_u-m_d}{m_s}\right)\right] 
 \lsim 9\times 10^{-12}~.\qquad
\eeqa
Any experimental limit on $BR(K_{L,S}\to\pi^0\nu\bar{\nu})$ below these values
carries a non--trivial dynamical information on the 
structure of the $s\to d\nu\bar{\nu}$ amplitude. 

In models where $K\to\pi \nu\bar{\nu}$ transitions are 
described by the effective Hamiltonian (\ref{Heff}), 
a measurement of $BR(K_{L}\to\pi^0\nu\bar{\nu})$ (or
$BR(K_{S}\to\pi^0\nu\bar{\nu})$) 
fixes $|\Imm C|$ (or $|\Real C|$), whereas 
$BR(K^+\to\pi^+\nu\bar{\nu})$ determines  $|C|$ (here, for simplicity, we are 
assuming lepton universality). 
Thus $C$ can be fixed up to a four--fold ambiguity.
In order to disentangle non SM effects it could be important to resolve
this ambiguity (see e.g. the discussion of \cite{Yuval}).
Even if present and foreseen facilities do not allow for this possibility,
we find it amusing to note that this could be done in principle in a very
high luminosity $\Phi$--factory, looking at  $K_{L,S}\to\pi^0\nu\bar{\nu}$
interference.  
Here, analogously to the double final state $|\pi^+\pi^-,\pi^0 \pi^0
\rangle$ analyzed for the measurement of $\epsilon'/\epsilon$,
one could study the time--difference distribution \cite{DIP}
of $|\phi\rangle \to |\pi^+\pi^-,\pi^0\nu\bar{\nu}\rangle$.
This is given by 
\beq
I(\pi^0\nu\bar{\nu},\pi^+\pi^-;t) \propto
\frac{e^{-\Gamma|t|}}{2\Gamma} \left\{ |\lambda_{\nu\bar{\nu}}|^2 
e^{-\frac{\Delta \Gamma}{2}t}
+  |\eta_{+-}|^2 e^{+\frac{\Delta \Gamma}{2}t} 
- 2\Real \left(\eta_{+-}^*\lambda_{\nu\bar{\nu}} e^{i\Delta m t}\right) \right\}~,
\eeq
where 
$t=t_{\pi^+\pi^-} - t_{\pi^0\nu\bar{\nu}}$,
$\Gamma=(\Gamma_S+\Gamma_L)/2$,  $\Delta m=m_L-m_S$,
$\Delta \Gamma=\Gamma_S-\Gamma_L$,
and 
\beq
\eta_{+-}=\frac{A(K_L\to\pi^+\pi^-)}{A(K_S\to\pi^+\pi^-)}\; , \; \; \;
\lambda_{\nu\bar{\nu}}=\frac{A(K_L\to \pi^0\nu\bar{\nu} 
)}{A(K_S\to \pi^0\nu\bar{\nu})} \; .
\eeq
Thus a measurement of the interference term would lead to an
unambiguous determination of the sign of $\lambda = i \; \Imm \; C/\Real \; 
C +{\cal O}(\epsilon)$. 

\section{Present Experimental Status and Prospects}

At present, the best published limit for the $K_L\to\pi^0\nu\bar{\nu}$ decay
is $5.8\times10^{-5}$ (90$\%$ C.L.), obtained by the FNAL experiment 
E799-I \cite{E799}. 
Recently, the KTEV Collaboration has presented a preliminary
result, giving an upper limit on the branching ratio of $1.8\times10^{-6}$ 
(90$\%$ C.L.) \cite{KamiP}.
The same Collaboration aims at reaching in 1999 a single
event sensitivity (that we will precisely define below) of $3\times
10^{-9}$.  

Sensitivities which should allow a positive
measurement of the branching ratio (assuming the Standard Model value) are
the goal of three dedicated experiments which have been recently
proposed. The first should run at the new 50 GeV high--intensity machine in
KEK \cite{KEKP};  
the second 
is the KAMI experiment at FNAL, essentially an upgraded continuation of 
the KTEV experiment \cite{KamiP}; 
finally, there is the BNL proposal \cite{BNLP},
whose approach 
is the closest to the one discussed in the present paper. In fact,
the BNL group 
proposes to execute the experiment on a micro--bunched, low--momentum 
($\sim$~700
MeV) kaon beam, with the purpose of measuring the momentum of the decaying 
$K_{L}$ with a time--of--flight technique, allowing the complete 
reconstruction of the decay kinematics. The advantages of this experimental
technique are similar to the ones discussed in the present 
paper, although the $\phi$--factory environment is free from the
uncertainties due to the presence of neutral halos 
in the kaon beam, typical of hadron machines.

However, the time scale for these experiments is such that the first
results will not be available before year 2003, at best. 

In the next section we will discuss the advantages of performing the
measurement at a $\phi$--factory, concentrating on the realistic case of the
KLOE experiment \cite{kloe1,kloe2,kloe3,kloe4,kloe5,kloe6}
at DA$\Phi$NE \cite{dafnr}. 
We will show that results as good as the 
one expected from KTEV 99 can be obtained in a relatively short time.

\section{$K_L\to\pi^0\nu\bar{\nu}$ at a $\phi$--factory}

At a $\phi$--factory, $\phi$(1020) mesons are produced at rest
by $e^+e^-$ collisions. Due to $C$--parity conservation, they decay  
into a \Kos--\Kol\ pair with a branching ratio of 34.1$\%$ \cite{PDG}.
By observing the \Kos\ decay into two charged pions, 
it is therefore possible to tag the presence of the \Kol\ 
moving in the opposite direction with a  $\sim$ 110 MeV/c momentum, 
determined by the $\phi$ decay kinematics. Therefore  the
complete reconstruction of the kinematics of the subsequent 
\Kol\ decay is allowed. 

Presently 
the newly built $\phi$--factory DA$\Phi$NE has begun commissioning  
in Frascati, with the peak luminosity of $5\times 10^{32}$ \lum.
At this luminosity, as many as 
10$^{10}$ correlated \Kos--\Kol\ pair per year can be
produced\footnote{Here and from now on,
following HEP convention, we define one physics year to be
equal to $10^{7}$~s.}.

The KLOE detector at DA$\Phi$NE, whose roll--in is expected
by mid 1998 \cite{kloe6}, has been 
designed and built with the main purpose of determining the 
$CP$--violating parameter $\epsp/\eps$
via the observation of the \Kol\
decays into two charged or two neutral pions. A very high
photon detection efficiency is one of the fundamental requirements 
for such a measurement; in particular great attention has been paid 
to the problem of minimizing the background from \Kol\ decays
into three neutral pions in which two photons escape detection
\cite{kloe1,kloe2}. 
For this reason KLOE is well suited also for 
the observation of the decay of interest in the present paper.
 
The detector consists of two main parts: a large cylindrical tracking 
chamber of 2 m radius and 3.7 m length, and a hermetic lead--scintillating 
fibers electromagnetic calorimeter (ECAL from now on). ECAL allows one
to detect photons with energy down to 20 MeV and to measure their
energy with a resolution of $\sigma_{E}=4.5\% \times\sqrt(E)$ ($E$ in
GeV). In addition, ECAL allows the determination of the entry--point
position of the photons with a resolution of 3 cm
and 1 cm for coordinates parallel and  perpendicular 
to the scintillating fibers, respectively.
Of great relevance is also the ability of the ECAL to determine the time
of the photon's passage with a resolution of 
$\sigma_{T}= 60 ps/\sqrt(E)$ ($E$ in GeV). 

In order to quantify the possible performance of such a 
detector, we have set up a simple Monte Carlo in which 
\Kol--\Kos\ pairs are generated from $\phi(1020)$ decays 
taking into account the correct energy and angular distributions; the \Kol\ 
is then allowed to decay into 2 $\pi^{0}$'s or 
$\pi^{0}\nu\bar{\nu}$, at a space point determined by its momentum and 
lifetime. 

We have concentrated our attention on the problem of the rejection of 
the $K_L\to\pi^0\pi^0$ background which is the key issue for the success
of the experiment (see section 4.3).  

We will see that, 
although the main physical ideas and some of the conclusions 
of our paper are a
generic consequence of the peculiar environment available at 
any $\phi$--factory, ultimately the detector's parameters, such as
geometrical acceptance and resolutions, 
become of decisive importance. We have therefore paid the maximum 
attention to the correct parametrization of such parameters trying, 
wherever possible, to check our conclusions with independent 
studies and official figures. 

\subsection{Description of the method}
In our code the KLOE detector is implemented as a cylinder of 4 m
diameter and 3.7 m length, hermetically closed at both ends by two
endcaps. In the following, the cylinder axis is defined as the $z$ axis.

\begin{figure}[t]
    \begin{center}
      \setlength{\unitlength}{1truecm}
       \begin{picture}(12,10)
       \epsfxsize 13.0 true  cm
       \epsfysize 13.0  true cm
       \epsffile{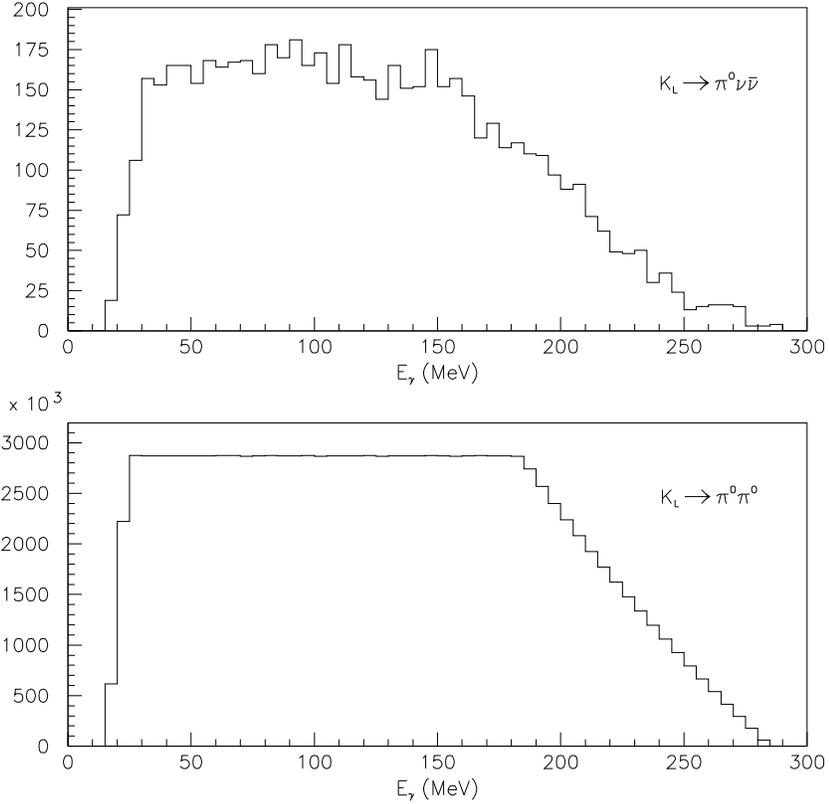}
       \end{picture} 
      \end{center}
    \caption{ Energy distribution for photons produced by 
$K_L\to\pi^0\nu\bar{\nu}$ (upper plot), and
$K_L\to\pi^0\pi^0$ events (lower plot) }
   \protect\label{fig1}
\end{figure}

One of the most important features of the KLOE detector is that it is
almost perfectly hermetic to photons.
There is however a small chance that a photon produced inside the
detector is lost. The causes of the losses are the following:
\begin{enumerate}
\item
There is a small region between the endcaps and the beam pipe where the
detector has a physical hole: this can be 
schematically modeled by two squares of
50 cm side, one for each endcap. 
\item
The beam pipe inside the detector and the wall of the drift chamber can
absorb photons. We have assigned a 2 \% probability of absorption to
photons intersecting the beam pipe or the internal walls of the drift
chamber, implemented as a cylinder of 20 cm radius, with
axis along the $z$ direction.
\item
Two sets of three low--$\beta$ quadrupoles are inserted along the 
beam pipe inside the detector. In order to detect the photons that 
would have been lost hitting these quadrupoles, the latter are covered 
by special calorimeters (QCAL). However, 
 the detection efficiency of QCAL is not
expected to exceed 90-95 $\%$ \cite{kloe7}:
 in the program it has been assigned a
90$\%$ efficiency, independent on the photon's energy. 
\item
Photon losses  in ECAL due to some detection 
inefficiency are always possible. In particular, low energy photons 
can be lost because of several effects, including sampling fluctuations, 
photonuclear reactions, reconstruction inefficiencies. A detailed study 
of these effects goes well beyond the scope of the present paper. We 
have parametrized them by assigning a 70$\%$ detection 
probability to 20 MeV photons, linearly increasing up to 100$\%$ at 
50 MeV. Photons with energy lower than 20 MeV were considered lost
both for ECAL and QCAL. 
The energy distribution of the photons produced by  
$K_L\to\pi^0\nu\bar{\nu}$ and $K_L\to\pi^0\pi^0$ decays is shown in 
Figure~\ref{fig1}. 
\end{enumerate}

\noindent
The program  computes the geometrical interception of the
photons produced by the decays which happen inside KLOE and the calorimeter,
and first of all decides whether the photons are lost or not.
In order to check the reliability of our simulation, we have generated
a sample of  $K_L\to\pi^0\pi^0\pi^0$ events and compared the number of
lost photons predicted by our program 
with the one predicted by the official KLOE 
Monte Carlo, GEANFI \cite{kloe1}.
In this comparison only geometrical effects were taken into 
account, i.e. the fourth source of photon losses discussed in the previous 
list was not considered. 
Decays happening inside a cylindrical fiducial volume defined by 
$-150< z <150$ and $40<R<180$ ($z$, $R$ in cm) were studied.   
Inside this fiducial volume, GEANFI
predicted a 0.83$\pm$0.02$\%$ loss for photons, while our 
simulation gave 1.44$\pm$0.01$\%$. 
The relative population of photons reaching the different parts
of the detector were in agreement; our simulations turned out to be only
slightly pessimistic in the prediction of photon losses on the beam--pipe
or on the internal wall of the drift chamber. 
This result gave us confidence that our
simplified Monte Carlo well reproduces  the main features of the KLOE 
detector, as far as photon detection is concerned. 

Once two photons reach the active part of the detector, they can be 
paired and their invariant mass can be computed. Here, detector's 
resolutions play a crucial role. 
For the energy and position resolutions of ECAL we have used the 
previously quoted figures. We have assigned a photon--energy
resolution of $\sigma_{E}=40\%$ to QCAL, independent of the photon's
energy, while keeping the same spatial resolution used for those hitting
ECAL.

The last piece of experimental information needed for the complete
reconstruction of the \Kol\ decay, is represented by the spatial 
coordinates of the decay vertex. 
Unique to the KLOE experiment is the method of determining it by time 
measurement. 
It has been shown 
that for events in which the \Kol\ decays into two neutral pions and the
\Kos\ decays into two charged ones, and where all the particles are
detected, this procedure allows a determination of the \Kol\ decay vertex
with uncertainties of order 0.6 cm on the three coordinates \cite{kloe2}.
Since in the events of interest for the present paper information is 
available only for two photons (instead of four), we have increased
this uncertainty to 1 cm. 

\subsection{Analysis of the $K_L \rightarrow \pi^0 \pi^0$ background}
We have generated two independent samples of events, consisting of
$10^{8}$  and $10^{4}$ \Kol's, respectively, out of which only those
decaying inside the fiducial volume defined by the conditions $|z| \lt$ 150
cm, and 40 cm$\lt R \lt$ 180 cm were studied. For the first sample, \Kol's
were forced to proceed through the channel  $K_L\to\pi^0\pi^0$, while for
the second through $K_L\to\pi^0\nu\bar{\nu}$. 
We then determined the fraction of events for which two and only two
photons were detected according to our simulation. These amount to 
0.23\% and 28\% for the first and the second sample, respectively.

\begin{figure}[t]
    \begin{center}
      \setlength{\unitlength}{1truecm}
       \begin{picture}(12,10)
       \epsfxsize 13.0 true  cm
       \epsfysize 13.0  true cm
       \epsffile{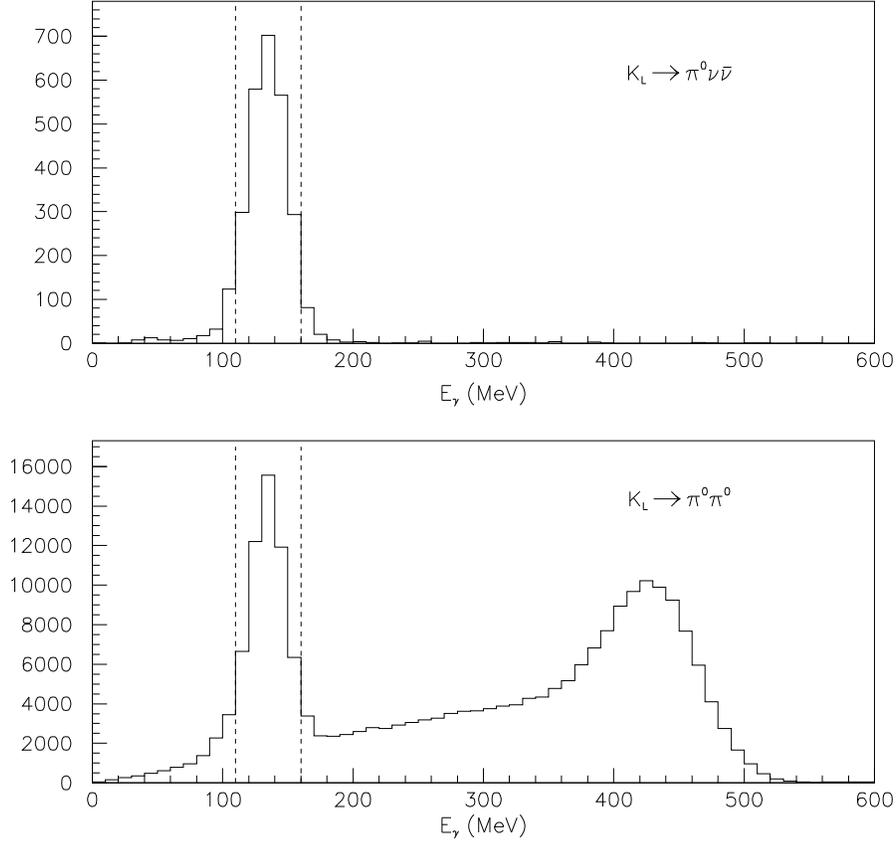}
       \end{picture} 
      \end{center}
    \caption{Two-photons invariant mass for events in which two and only
 two $\gamma$'s are detected. Upper plot: $K_L\to\pi^0\nu\bar{\nu}$ decays.
    Lower plot: $K_L\to\pi^0\pi^0$ events. The dashed lines denote
    the mass window used in the analysis.}
    \protect\label{fig2}
\end{figure}

\begin{figure}[t]
    \begin{center}
      \setlength{\unitlength}{1truecm}
       \begin{picture}(12,10)
       \epsfxsize 13.0 true  cm
       \epsfysize 13.0  true cm
       \epsffile{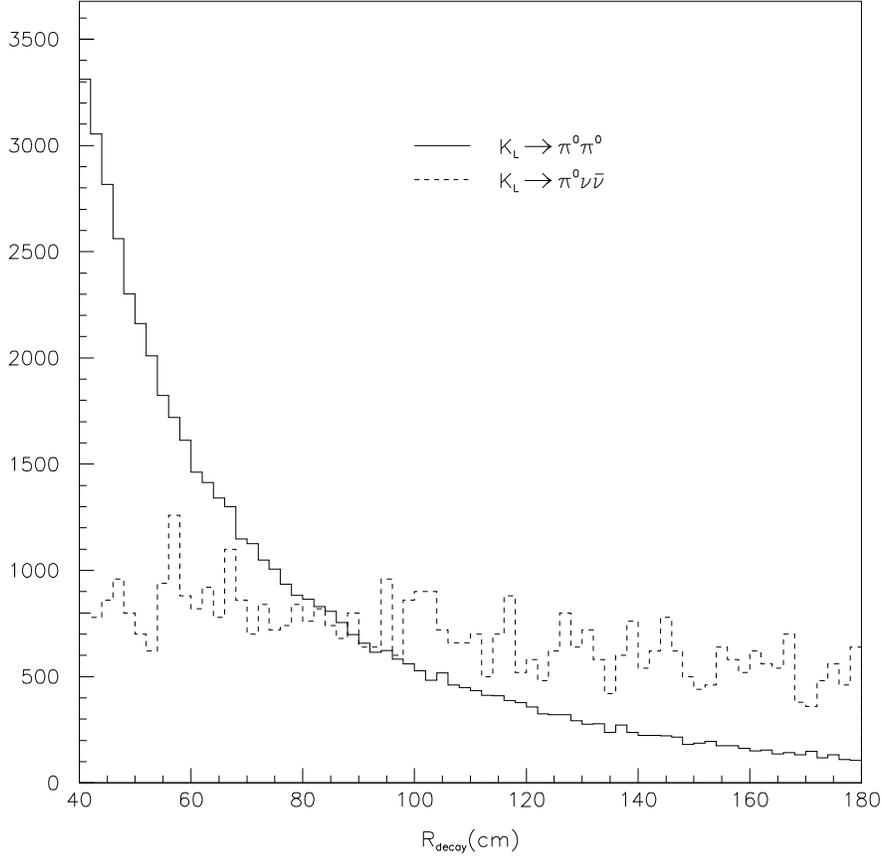}
       \end{picture} 
      \end{center}
    \caption{Decay radius for two-photons events from
     $K_L\to\pi^0\nu\bar{\nu}$ (dashed line) and $K_L\to\pi^0\pi^0$ 
     (solid histogram) decays. As can be noticed, background events are 
      concentrated mainly at positions close to the beam line. 
      The two histograms are not on scale.}
    \protect\label{fig3}
\end{figure}

\begin{figure}[t]
    \begin{center}
      \setlength{\unitlength}{1truecm}
       \begin{picture}(12,10)
       \epsfxsize 13.0 true  cm
       \epsfysize 13.0  true cm
       \epsffile{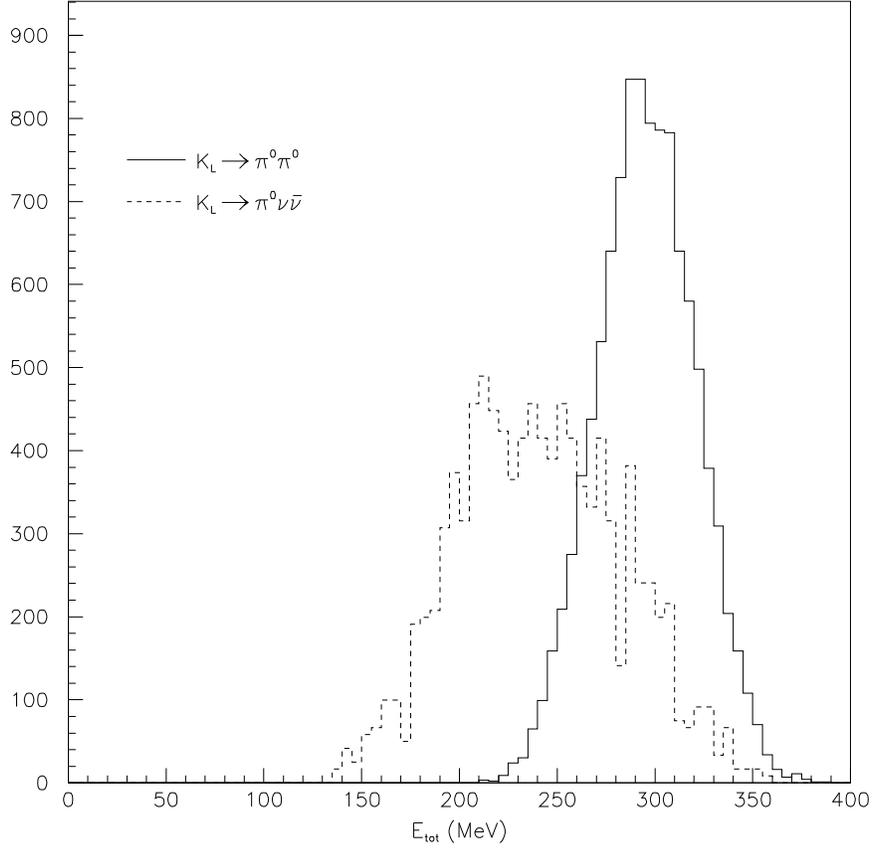}
       \end{picture} 
      \end{center}
    \caption{Total reconstructed energy for two-photons events from
    $K_L\to\pi^0\nu\bar{\nu}$ (dashed line) and $K_L\to\pi^0\pi^0$
    (solid histogram) decays.} 
    \protect\label{fig4}
\end{figure}

\begin{figure}[t]
    \begin{center}
      \setlength{\unitlength}{1truecm}
       \begin{picture}(12,10)
       \epsfxsize 13.0 true  cm
       \epsfysize 13.0  true cm
       \epsffile{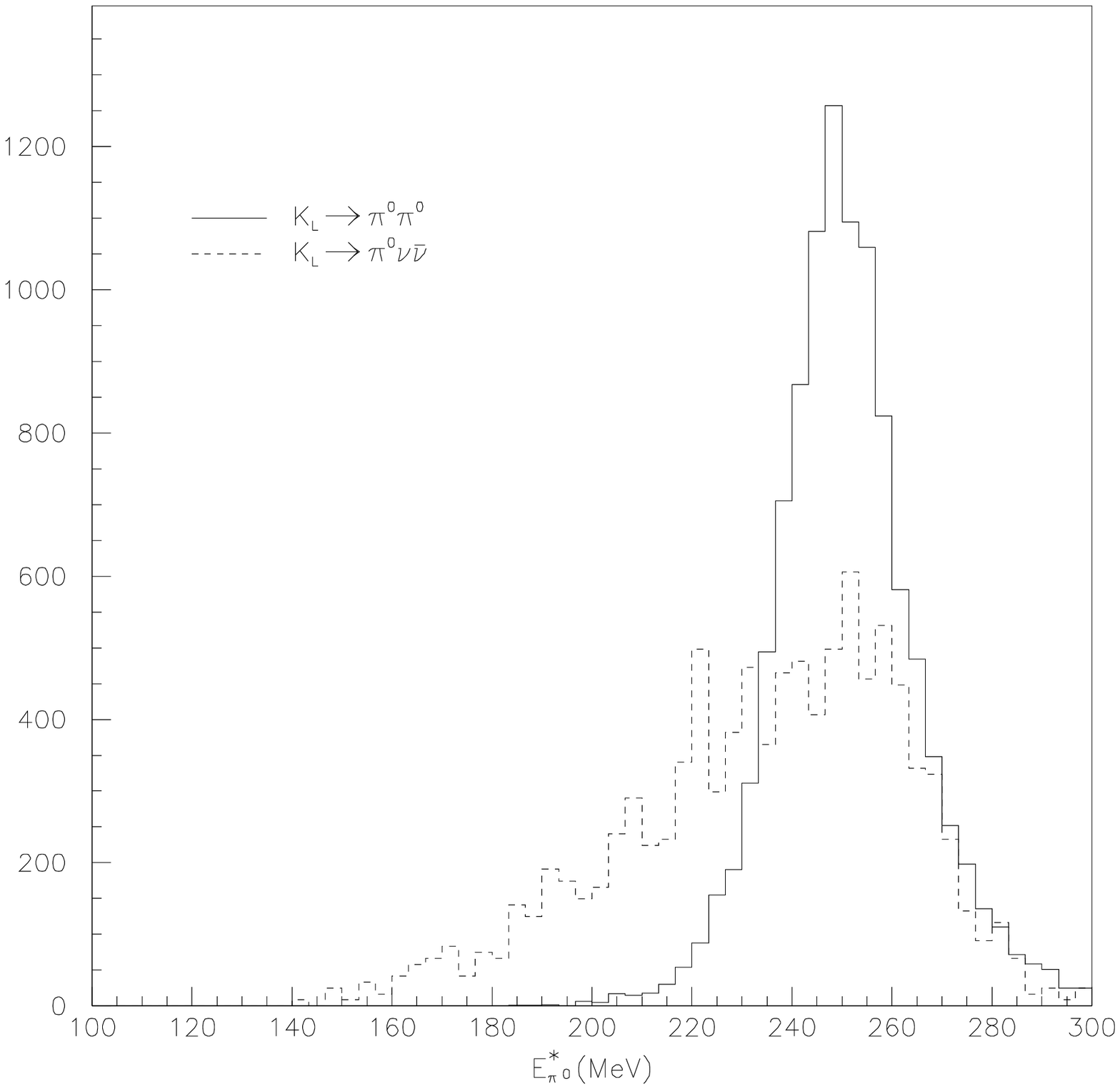}
       \end{picture} 
      \end{center}
    \caption{ Reconstructed $\pi^{0}$ energy in the \Kol\ rest frame
      for two-photon events from    
    $K_L\to\pi^0\nu\bar{\nu}$ (dashed line) and $K_L\to\pi^0\pi^0$
    (solid histogram) decays.} 
    \protect\label{fig5}
\end{figure}

On this sample of events with only two detected photons we have made the
following analyses:
\begin{enumerate}
\item
We have studied the distribution of the reconstructed
two--photon invariant mass, $M^R_{\gamma\gamma}$, 
after resolution effects are taken into account (Figure~\ref{fig2}). 
In the $K_L\to\pi^0\pi^0$ case more than 70\%
of the  events are due to odd--pairings 
(i.e. the two photons come from two different $\pi^{0}$'s),
and can be easily removed by a cut on $M^R_{\gamma\gamma}$.
Imposing $|M^R_{\gamma\gamma} - M_{\pi^0}| < 25$ MeV
only 23\% of the $K_L\to\pi^0\pi^0$ background survives whereas 
the signal efficiency is 87\%.
\item
We then analyzed the distribution of the decay positions of the 
two samples, once the above cut on $M^R_{\gamma\gamma}$ is applied. 
Since the dead zones are not
uniformly distributed inside the detector, the background
distribution is expected to be peaked around the beam line.
Figure~\ref{fig3} shows the amount of two photon 
events as a function of the decay radius,  $R_{\rm decay}$, for both signal 
and background. It can be seen that a cut on the minimum allowed decay 
radius can increase significantly the signal/background ratio, at 
the price of somewhat lowering the signal detection efficiency. 
For instance, by choosing events for which 
$R_{\rm min} = 100$~cm~$< R_{\rm decay} < 180$~cm,  
gives a signal efficiency of 50\% and reduces 
the background to 20\%. 
The combined cuts on $M^R_{\gamma\gamma}$ and $R_{\rm min}$,
together with the two--photon requirement,
leads to an overall background rejection of $\sim 10^{-4}$ and
a 12\% signal efficiency.
\item
The use of the previously defined acceptance volume has the other advantage
of providing a new powerful handle for background rejection.  Since
dead zones are concentrated {\it backwards} with respect to the \Kol\
flight direction, momentum conservation implies that lost photons from 
$K_L\to\pi^0\pi^0$ decays are mostly low energy ones in the laboratory
frame. Consequently the distribution of the total reconstructed energy 
for two--photon events from $K_L\to\pi^0\pi^0$ decays has to be displaced
towards high values, as shown by Figure~\ref{fig4}. Conversely, photons
from signal events may have lower energies because a significant part of
the total energy can be carried away by the two neutrinos. A cut around 
$E_{\rm tot}=0.22$~GeV leads to an additional $10^{-3}$ suppression
of the $K_L\to\pi^0\pi^0$ events.
\newline
The power of this method rests on two facts; firstly, one knows {\it 
a priori} what is the total available energy in the decay, since the
\Kol\ beam is monochromatic. Secondly, as stated above, the detector's 
dead zone are concentrated in a well defined region with respect to the 
\Kol\ flight direction (which in turn is determined by the 
$\phi$ decay angular distribution). 
\item
The other possible strategy for background rejection rests
on the possibility of reconstructing the $\pi^{0}$ energy 
in the \Kol\ rest frame ($E^{\ast}_{\pi^{0}}$). In this frame, $\pi^{0}$'s 
from $K_L\to\pi^0\pi^0$ transitions are monochromatic with 
$E^{\ast}_{\pi^{0}}=M_{K^{0}}/2$, while for signal events 
$E^{\ast}_{\pi^{0}}$ ranges from $M_{\pi^0}$ to 
$E_\pi^{\rm max}$ defined in (\ref{energyrange}).
 However, due to the finite detector's resolution,  
the two distributions overlap as shown in 
Figure~\ref{fig5}. A cut around $E^{\ast}_{\pi^{0}}=0.2$~GeV
leads again to a $10^{-3}$ suppression
of the residual $K_L\to\pi^0\pi^0$ events.
\end{enumerate}

\noindent
We have found that the two strategies, (i.e. cutting on 
$E_{\rm tot}$ or cutting on $E^{\ast}_{\pi^{0}}$) are
very much correlated. Once either of the two cuts has been 
applied, the other is almost totally ineffective in decreasing 
the background. This can be understood since 
the detection of higher energy photons, which is 
an obvious requirement for the success of the first
strategy, improves also the precision with which $E^{\ast}_{\pi^{0}}$ 
is reconstructed, since the ECAL energy resolution 
scales as $\sqrt{E}$. In other words, background events 
in the low--energy tails of both $E^{\ast}_{\pi^{0}}$ 
and $E_{\rm tot}$ distributions are strongly correlated.
Furthermore, since the \Kol\ momentum is small, 
it is also clear that the low $E_{\rm tot}$
region contains mainly signal events
with small $E^{\ast}_{\pi^{0}}$.

\subsection{Other background sources}

Although the branching ratio of the $K_L\to 3 \pi^0$ is $\sim$250
times larger than the $K_L\to\pi^0\pi^0$ one, 
the probability of detecting only one $\pi^0$ in the former kind of events
is much suppressed with respect to the latter. 
Our Monte Carlo predicts a probability of about 10$^{-8}$ of observing
only two photons with $100 \; \mbox{cm} < R_{\rm decay} < 180 \; \mbox{cm}$
from $K_L\to 3 \pi^0$. We then estimate that applying the same cuts
as in the $K_L\to\pi^0\pi^0$ case this background can be reduced to the
level of about 10$^{-10}$.

A four--photon final state is produced also by the $K_L\to\pi^0\gamma\gamma$
transitions. However, the branching ratio of these events is a factor 
of $\sim$ 500 lower than that of $K_L\to\pi^0\pi^0$. Moreover,
of the four photons in the final state only two belong to a
$\pi^0$, so that the cut on $M_{\gamma\gamma}$ is expected to work
better than in the $K_L\to\pi^0\pi^0$ case. Since the rejection of the 
latter is at the $10^{-4}$ level before any cut $E_{\rm tot}$ 
or $E^*_{\pi^0}$, the $K_L\to\pi^0\gamma\gamma$ background does not 
represent a problem, at least for a search on 
$K_L\to\pi^0\nu\bar{\nu}$ above $10^{-10}$.
  
Finally, all the \Kol\ decays involving one $\pi^0$ and two charged particles
should be easily rejected, since
the KLOE drift chamber is able to detect with very high efficiency,
within the fiducial volume, the presence of charged particles \cite{kloe3}. 

\subsection{Discussion of the results}

\begin{table}[t]
  \begin{center}
    \leavevmode
    \begin{tabular}{|c||c|c||c|l|l|}\hline
       & & & & & \\
       & Energy cut (GeV) & $R_{\rm min}$(cm)  & $\epsilon$(\%) & ~~~~SES &
       ~~~ BR1 \\
       & & & & & \\
 \hline
       & & & & & \\
       & $E^*_{\pi^0}<0.20$ & 100  & 1.1
       & $6\times10^{-9}$   & $6\times10^{-9}$ \\    
KLOE   & $E_{\rm tot} < 0.22$   & 100  & 2.8
       & $1\times10^{-9}$   & $2\times10^{-9}$ \\    
\ $\int{\cal L}= 10^{40}$~cm$^{-2}$\       & $E_{\rm tot} < 0.22$   & 90   & 3.1
       & $3\times10^{-9}$   & $2\times10^{-9}$ \\ 
       & $E_{\rm tot} < 0.21$   & 90   & 2.3
       & $5\times10^{-10}$  & $3\times10^{-9}$ \\ 
       & & & & & \\
     \hline
     \hline
       & & & & & \\
Ideal & $E^*_{\pi^0}<0.22$ & 100  & 2.8
       & $< 10^{-10}$       & $2\times10^{-10}$ \\    
 detector& $E_{\rm tot} < 0.24$   & 100  & 5.7
       & $1\times10^{-10}$  & $1\times10^{-10}$ \\    
\ $\int{\cal L}= 10^{41}$~cm$^{-2}$\  & $E_{\rm tot} < 0.23$   & 100 &  4.1
       & $< 10^{-10}$       & $2\times10^{-10}$ \\  
       & & & & & \\ \hline   
    \end{tabular}
    \caption{Efficiency, single event sensitivity (SES) and branching ratio
      of the signal that would yield one event (BR1) corresponding to
      different cuts in two typical situations: (1) the KLOE detector, with
      features as described in the text and for an integrated luminosity
      of $10^{40}$~\ilum; 
      (2) our definition of an ideal detector, with improved
      calorimeters as described in the text and for an integrated luminosity 
      of $10^{41}$~\ilum.}  
    \label{tab1}
  \end{center}
\end{table}

Table~\ref{tab1} summarizes our results. The case of KLOE
and DA$\Phi$NE corresponds to the values in the upper part of the table. In
the first two columns various combinations for the values
of the cuts on $E_{\rm tot}$ (or $E^{\ast}_{\pi^{0}}$) and on $R_{\rm min}$, 
respectively, are listed. The efficiency obtained applying these 
cuts on signal events is shown in the third column. In the fourth column
for any given set of cut values the Single Event Sensitivity (SES) is
given, defined as the value of the $K_L\to\pi^0\nu\bar{\nu}$ branching
ratio for which the expected number of signal
events equals that of background ones.  
Finally, in the last column, the value of the $K_L\to\pi^0\nu\bar{\nu}$
branching ratio for which one event is expected after two years of data 
taking at the luminosity of $5\times$10$^{32}$ \lum\ (BR1) is 
shown.

Note that there are two important effects which determine all  
the quoted signal efficiencies: the $\sim$~67$\%$ 
branching ratio of the $K_S\to\pi^+\pi^-$ decay which is used for tagging
purposes, 
and the fact that about 2/3 of the produced \Kol's do not decay 
before reaching ECAL. Therefore, 
once a given set of cut values is chosen, the minimum obtainable value
for BR1 is ultimately determined by the luminosity that can be delivered
by the machine. 
In the case of DA$\Phi$NE one can hope to reach luminosities up to a factor
$\sim$ 2 larger than the nominal value, without modifying the hardware
set--up of the machine. On the other hand, a good experiment
should aim at reaching the minimum possible BR1, while 
keeping SES $\lt$ BR1. A reasonable figure of merit 
 for any given set of cut values is therefore the ratio 
R$_{\rm mer}=$BR1/SES, which should be kept $\geq$ 1; 
it is seen that, in the case of KLOE, branching ratios 
of order $10^{-9}$ with figures of merit R$_{\rm mer} \sim 1-2 $
can be obtained, at best.
This would already be a competitive measurement for 
several years to come. 

Interestingly enough, there is not much space for possible improvements
with the KLOE detector, since the benefits of a higher luminosity, 
which can decrease BR1, would be spoiled by the obtainable SES's,
i.e. by the presence of an irreducible amount of background events. 
Although our analysis cannot be considered
exhaustive and the possibility for a wiser and more effective strategy 
with KLOE and DA$\Phi$NE can always be considered, we believe that
significant improvements on these figures can be obtained only by combining
a more  efficient detector with a higher luminosity accelerator. 
In particular, our simulation showed that the main problems 
of the present detector arise from losses of soft photons 
($E<50$ MeV) and from the spread in the energy resolution.
For this reason, we have considered
the possibility of one year of running
at a luminosity of 1$\times$10$^{34}$ \lum, with a detector with the same
geometry as KLOE but a better resolution, given by the following
parameters:  $\sigma_{E}=2\% \times\sqrt(E)$  
and 100\% efficiency for photons with $E>20$ MeV, both for QCAL and ECAL. 
It is seen that one not only improves in the reachable BR1 (thanks mainly to
the higher luminosity), but also becomes  more efficient in the rejection
of the background, decreasing substantially the SES. 
Unfortunately, to our understanding,
both the machine and the detector's parameters used in this case
are not reachable in the next few years.

\begin{figure}[t]
    \begin{center}
      \setlength{\unitlength}{1truecm}
       \begin{picture}(12,8)
       \epsfxsize 10.0 true  cm
       \epsfysize 12.0  true cm
       \epsffile{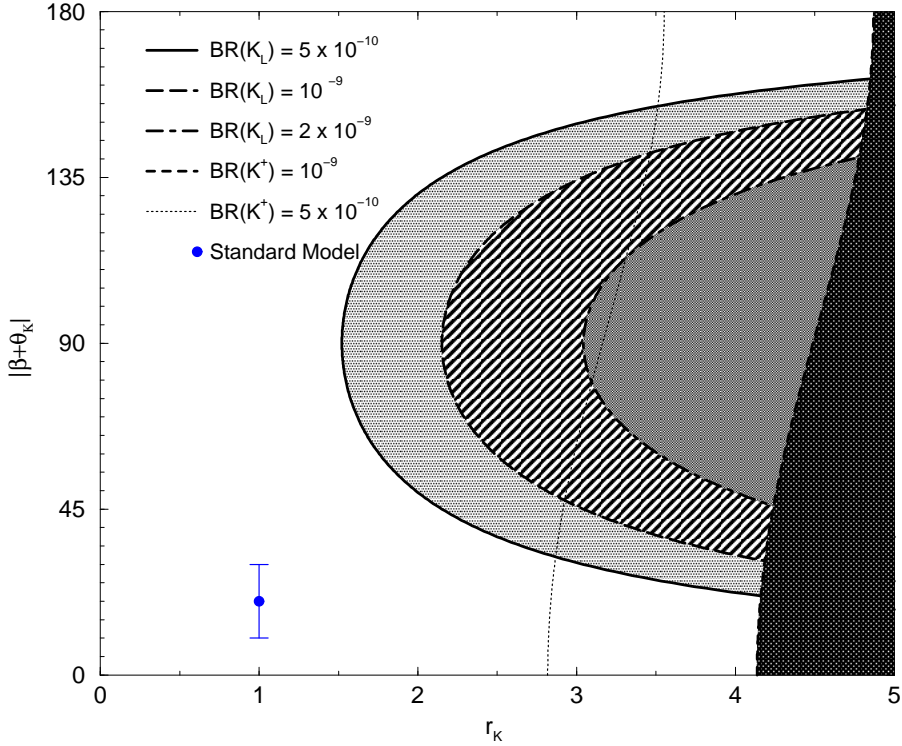}
       \end{picture} 
      \end{center}
    \caption{Combined information coming from 
limits/observations of $K_L\to\pi^0\nu\bar{\nu}$
and $K^+\to\pi^0\nu\bar{\nu}$ around the $10^{-9}$ level.
The parameters $r_K$ and $\theta_K$, describing New--Physics
effects in the $s\to d\nu\bar{\nu}$ amplitude are 
defined in Eq.~(\protect\ref{Clbis}).
The curves have been obtained assuming 
the central values of $|\lambda_{t,c}|$, $X^l_{NL}$ and
$X(m_t^2/M_W^2)$, as reported in \protect\cite{BBL}. The SM scenario 
is recovered for $r_K=1$ and $\theta_K=0$.}
    \protect\label{fig6}
\end{figure}

Figure~\ref{fig6} summarizes the physical information coming from an
observation/search for both the neutral-- and charged--kaon decay of
interest here. On the two axes we have the two  
parameters $r_K$ and $\theta_K$ (in fact the modulus of the sum of this
phase plus the SM phase $\beta$), defined in Eq. (\ref{Clbis}).
To each value of the two branching ratios $BR(K_L)$ and $BR(K^+)$
there corresponds a different curve in the $(r_K,\;\theta_K)$ plane, as
shown in Fig.~\ref{fig6}.
A positive measurement of both branching ratios would allow one
to pin down
(modulo a two--fold ambiguity) the value of both parameters.
Upper limits on the branching ratios only allow the exclusion of regions in 
that plane. The rightmost curve and shaded area on the figure 
correspond to the reference value of $BR(K^+)\leq 10^{-9}$, 
which is close to the current bound coming from 
the E787 experiment \cite{BNL}.
It is clear that the Standard Model value of the two parameters is far away
from that curve, and that there is a very large region in parameter space
to be explored. The curves corresponding to $BR(K_L)=(2, \;
1, \; 0.5) \times 10^{-9}$ show the possible improvements that a search for 
this decay in the following few years at a $\phi$--factory could bring.
In particular, the comparison to the curve corresponding to $BR(K^+)=5
\times 10^{-10}$ shows very clearly that the two measurements/searches are
complementary to each other: the $BR(K^+)$ ($BR(K_L)$) measurement
strongly constrains the value of $r_K$ ($\theta_K$), leaving $\theta_K$
($r_K$) practically undetermined. Even if $BR(K^+)$ was
measured with rather small uncertainties, and found in agreement with the
SM value, there would still be the possibility to have a 
{\it new--physics  phase} 
$\theta_K$ very different from zero, and only the $BR(K_L)$
measurement could exclude this interesting scenario.

\section{Conclusions}

The observation of the $K_L\to\pi^0\nu\bar{\nu}$ transition is of the
utmost relevance, since it provides very clean information on one of the
less known CKM matrix elements, and also because it could signal the
presence of new physics beyond the Standard Model.
The experimental challenge to perform such a measurement is a very
difficult one, because of the very low expected branching ratio ($\sim$
3$\times$10$^{-11}$ in the Standard Model), and the presence of
copious sources of background events which could fake the signal. 

At present, there are three proposals for experiments which claim to be able 
to measure the SM branching ratio with a $\sim$10$\%$ precision. None of
these, however, will produce results for several years to come. 
On the other hand, we have seen that any measurement which can improve on
the phenomenological limit $5 \times 10^{-9}$  carries a non--trivial
dynamical information on the structure of the $s\to d\nu\bar{\nu}$
amplitude, which at present is very poorly known, and would therefore
constrain the parameter space of possible extensions of the SM.

We have argued that with the KLOE detector at DA$\Phi$NE it is possible to
lower considerably the present experimental upper bound within a few
years. Our main point is that the $\phi$--factory environment is naturally
well suited for the solution of the most difficult experimental problem,
i.e. the rejection of the $K_L\to\pi^0\pi^0$ background.
Moreover, we have shown that the particular geometry of KLOE, a detector 
which was conceived and built to minimize the inefficiency in detecting
photons, provides excellent possibilities to discriminate
between signal and background events. 
With the present facility, one can reach a sensitivity to branching
ratios of 10$^{-9}$ or lower, in some years of running.  
This does not allow a positive observation of the
Standard Model $K_L\to\pi^0\nu\bar{\nu}$ transition: only a serious
improvement in the delivered luminosity and in the detector's parameters
would allow one to reach this ambitious goal.
On the other hand Fig.~\ref{fig6} very clearly shows that KLOE has a chance 
to provide unique and invaluable information in excluding possible
deviations from the SM in the value of the phase $\theta_K$.

The results of the present paper are meant mainly as a first,
conservative estimate: only a dedicated detailed study on systematic
effects could yield precise numbers on the sensitivity the 
KLOE detector could reach for such a decay. 
Our main aim was to show that this study is worthwhile and that an effort
in this direction should be seriously taken into consideration. In this
respect, we find it particularly relevant that this measurement does not
require any modification in the data taking plans of KLOE. 
Moreover it is obvious that a detailed study of all the 
effects which may affect photon detection in KLOE 
is of the highest importance also with respect to $\epsp/\eps$ studies,
which are the main concerns of the Collaboration. 

\section*{Acknowledgments}
It is a pleasure to thank G.~Buchalla, P.~Franzini, Y.~Grossman,
J.~Lee--Franzini, G.~Vi\-gno\-la and M.~Worah for interesting discussions
and comments on the manuscript. We are indebted to E.~Santovetti
for providing us with files containing the results of the official
KLOE Monte Carlo.

 
\end{document}